\pdfoutput=1

\documentclass[11pt]{article}
\usepackage{jheppub}

\usepackage{amsmath}
\usepackage{amssymb}
\usepackage{graphicx,color,slashed,braket}
\usepackage{bbm}
\usepackage{ifpdf}
\usepackage{comment}

\newcommand{\cF}{\mathcal{F}}

\newcommand{\be}{\begin{equation}}
\newcommand{\ee}{\end{equation}}
\newcommand{\ba}{\begin{eqnarray}}
\newcommand{\ea}{\end{eqnarray}}

\newcommand{\lp}{\left(}
\newcommand{\rp}{\right)}
\newcommand{\ls}{\left[}
\newcommand{\rs}{\right]}

\newcommand{\vp}{\varphi}

\newcommand{\w}{\wedge}

\newcommand{\N}{\mathcal{N}}

\def\rmi{{\rm i}}
\def\rme{{\rm e}}

\def\Re{\mathop{\rm Re}\nolimits}
\def\Im{\mathop{\rm Im}\nolimits}

\def\K{{K\"{a}hler} }

\def\jb{{\bar \jmath}}

\title{Non-supersymmetric branes}

\author[a]{Niccol\`o Cribiori,}
\author[a]{Christoph Roupec,}
\author[b]{Magnus Tournoy,}
\author[b]{Antoine Van Proeyen,}
\author[c, a]{and Timm Wrase}

\affiliation[a]{Institute for Theoretical Physics, TU Wien,\\
Wiedner Hauptstrasse 8-10/136, A-1040 Vienna, Austria}
\affiliation[b]{Instituut voor Theoretische Fysica, KU Leuven,\\ Celestijnenlaan 200D, B-3001 Leuven,
Belgium}
\affiliation[c]{Department of Physics, Lehigh University,\\
16 Memorial Drive East, Bethlehem, PA, 18018}

\emailAdd{niccolo.cribiori@tuwien.ac.at}
\emailAdd{christoph.roupec@tuwien.ac.at}
\emailAdd{magnus.tournoy@kuleuven.be}
\emailAdd{Antoine.VanProeyen@fys.kuleuven.be}
\emailAdd{timm.wrase@lehigh.edu}

\notoc

\abstract{

\noindent
We discuss how to incorporate non-supersymmetric branes in compactifications of type II string theories. We particularly focus on flux compactifications on $SU(3)\times SU(3)$ structure manifolds to four dimensions, so that a linear $\N=1$ supersymmetry is spontaneously broken by spacetime filling Dp-branes. Anti-Dp-branes are a very special subset of such branes but our analysis is generic. We show that the backreaction of non-supersymmetric branes can be incorporated into the standard 4d $\N=1$ supergravity by including a nilpotent chiral multiplet. Supersymmetry in such setups is always spontaneously broken and non-linearly realized. In particular this means that, contrary to what was previously thought, brane supersymmetry breaking cannot be simply described by a D-term in 4d $\N=1$ supergravity theories.
}

\begin{document}

\maketitle

\newpage

\section{Introduction}\label{sec:introduction}
The fact that our universe is not supersymmetric clearly indicates the importance of studying compactifications of string theory where supersymmetry is broken in four dimensions. Beside fluxes, also local sources in string theory, like Dp-branes and Op-planes, generically break some amount of the original supersymmetry. By choosing appropriately the microscopic ingredients, it is possible to construct string compactifications in which the breaking of supersymmetry in four dimensions is complete, but spontaneous. When supersymmetry is spontaneously broken, it can then be non-linearly realized. In this large class of model, particularly interesting are those in which there is no order parameter capable to restore linear supersymmetry, namely supersymmetry is intrinsically non-linear. These models involve non-supersymmetric Dp-branes, as we will show in the present work.

In view of its connections to string theory, the study of non-linear supersymmetry has been revived recently, building on the early work \cite{Kachru:2002gs}, and especially in relation with the goldstino on an anti-D3-brane \cite{Kallosh:2014wsa,Bergshoeff:2015jxa,Kallosh:2015nia,Garcia-Etxebarria:2015lif,Vercnocke:2016fbt,Kallosh:2016aep,Aalsma:2018pll}. Indeed, the proper framework in which to insert all of these constructions is that of brane supersymmetry breaking \cite{Sugimoto:1999tx, Antoniadis:1999xk, Angelantonj:1999jh, Aldazabal:1999jr, Angelantonj:1999ms,Dudas:2000nv, Pradisi:2001yv} (see \cite{Mourad:2017rrl} for a recent review), in which supersymmetry is broken at the string scale by the presence of non-mutually BPS objects in the vacuum and it cannot be restored below that scale.

In this work, we study non-supersymmetric Dp-branes in string compactifications. In particular, we concentrate on the very broad class of flux compactifications of type II string theory on $SU(3)\times SU(3)$ structure manifolds to four-dimensional theories preserving a linear $\N=1$ supersymmetry. This is then spontaneously broken and non-linearly realized due to the presence of spacetime filling Dp-branes. The first appearance of such kind of branes in flux compactifications on $SU(3)\times SU(3)$  structure manifold is in \cite{Lust:2008zd}. It is known that building a correct dictionary between the ten-dimensional and the four-dimensional setup can be very subtle. In this regard, we correct a long-standing misconception in the literature where such non-supersymmetric Dp-branes in four-dimensional $\N=1$ supergravity have been described using D-terms.\footnote{Note, that this does not mean that any of the related older papers contains wrong results. Usually the D-brane action is properly reduced and used in the analysis. It is, however, often incorrectly stated that in the setting of 4d $\N=1$ supergravity the terms that result from the reduction of the non-supersymmetric D-brane action can be written in terms of a D-term.} We here provide consistent four-dimensional supersymmetric actions which reproduce precisely their ten-dimensional counterparts. In order to obtain such general expressions, we have to set the model-dependent world volume fields to zero and include only the universal Goldstino field that lives on the Dp-brane world volume. Then we can give a precise expression for the scalar potential that contains the couplings of all closed string moduli to the Dp-branes and that is manifestly invariant under non-linear supersymmetry.

We start by considering spacetime filling Dp-branes in flat space. This setup can be considered as a simple toy model in which the main differences between four-dimensional linear and non-linear supersymmetry can be understood from the brane perspective. In section \ref{sec:linvsnonlin} we show how Dp-branes at angles with respect to an Op-plane (or equivalently Dp-branes with world volume fluxes) lead to a spontaneous breaking and non-linear realization of supersymmetry, governed by the string scale. For this reason, we generically call these objects \emph{non-supersymmetric branes}.

Then, we show that, once supersymmetry is in the non-linear phase and the order parameter, independently of the moduli contains a constant piece, then there does not exist any field redefinition that absorbs the order parameter and brings you to linear supersymmetry. This means that  string compactifications in which supersymmetry is broken by local sources, such that the supersymmetry breaking scale is related to the string scale, cannot be described using the language of standard linear supersymmetry or supergravity. This corrects a misconception in the literature where people tried to use the familiar language of linear supergravity and in particular D-terms to describe such non-supersymmetric branes.

In section \ref{sec:nonsusybranes} we consider non-supersymmetric Dp-branes in generic backgrounds with $SU(3)\times SU(3)$-structure, following \cite{Kallosh:2018nrk}. While the case of a D3 brane is somehow peculiar, since it cannot be placed at arbitrary angles, we provide a unified description of the $p=5,6,7,9$ cases. Then, we give a general recipe to recast consistently their couplings in a ${\cal N}=1$ four-dimensional supersymmetric language, specifying a K\"ahler potential and a superpotential. In other words, we build a precise dictionary between ten-dimensional and four-dimensional quantities. The formalism of constrained multiplets is conveniently adopted for this purpose. The anti-D3-brane and intersecting D6-branes are discussed in more detail in specific examples before we give the general answer.

\section{Linear and non-linear supersymmetry}\label{sec:linvsnonlin}
In this section we recall some basic facts about Dp-branes and how they spontaneously break half of the supersymmetry in flat space \cite{Polchinski:1998rr}. We discuss branes at angles and the difference between anti-Dp-branes and Dp-branes. We also review supersymmetry breaking by world volume fluxes, which is T-dual to branes intersecting at angles \cite{Johnson:2003gi}. Then we show that SUSY breaking by branes necessarily requires a description in terms of non-linear supersymmetry. Lastly, we discuss the restoration of supersymmetry at high energies, exemplified by some explicit string theory setups.

\subsection{D-branes in flat space}
Let us start by studying Dp-branes in type II supergravity in flat space $\mathbb{R}^{9,1}$. This allows us to easily review a few simple facts \cite{Cederwall:1996pv, Aganagic:1996pe, Cederwall:1996ri, Bergshoeff:1996tu, Aganagic:1996nn}. The type II supergravity background preserves 32 supercharges that can be conveniently packaged into two Majorana-Weyl spinors $\epsilon_1$ and $\epsilon_2$.\footnote{These spinors have opposite (the same) chirality in type IIA (IIB). This distinction will not be relevant for us and we will treat type IIA and type IIB at the same time whenever possible.} Adding a Dp-brane along the directions $x^0, x^1, \ldots, x^p$ breaks half of the supersymmetry spontaneously \cite{Polchinski:1998rr}. The 16 supercharges that act linearly on the Dp-brane world volume fields satisfy
\be\label{eq:Dp}
\epsilon_1 = \Gamma_{01\ldots p} \epsilon_2 \equiv \Gamma_{\rm Dp} \epsilon_2\,,
\ee
where $\Gamma_{01\ldots p} = \Gamma_0 \Gamma_1\ldots \Gamma_p$ with $\Gamma_M$ the flat $\Gamma $-matrices of $D=10$. The 16 supercharges that satisfy $\epsilon_1 = - \Gamma_{\rm Dp} \epsilon_2$ are spontaneously broken and non-linearly realized on the world volume fields of the Dp-brane.

For an anti-Dp-brane things are exactly opposite and the linearly realized and unbroken 16 supercharges satisfy
\be
\epsilon_1 = -\Gamma_{01\ldots p} \epsilon_2 = -\Gamma_{\rm Dp} \epsilon_2 \equiv \Gamma_{\overline{\rm Dp}}\, \epsilon_2\,,
\ee
while the 16 non-linearly realized supercharges satisfy $\epsilon_1 = \Gamma_{\rm Dp} \epsilon_2 \equiv -\Gamma_{\overline{\rm Dp}}\, \epsilon_2$.

More generically we find for a Dp-brane that extends along $x^0, x^1, \ldots, x^{p-1}$ and intersects the $x^p$-axis with angle $\vp$ in the $(x^p, x^{p+1})$-plane, that the 16 linearly realized supercharges are solutions to
\be\label{eq:Dpangle}
\epsilon_1 = \lp \cos(\vp) \Gamma_{01\ldots p}+\sin(\vp) \Gamma_{01\ldots (p-1)(p+1)}\rp \epsilon_2 \equiv \Gamma_{\rm Dp}(\vp) \epsilon_2\,.
\ee
We then find trivially $\Gamma_{\rm Dp}(0) = \Gamma_{\rm Dp}$ and $\Gamma_{\rm Dp}(\pi) = \Gamma_{\overline{\rm Dp}}$. This simply means that an anti-Dp-brane is the same as a Dp-brane with opposite orientation.

Note that in the above discussion the entire separation into branes, anti-branes and branes at angles is rather artificial and not physically meaningful since we can always change (rotate) our coordinate system. So, the three separate cases above are all physically equivalent since we started with a maximally supersymmetric background. However, if the background breaks some of the supersymmetry, then these cases are not necessarily equivalent anymore.

Let us consider again the flat space case and do an orientifold projection that projects out half of the 16 supersymmetries. In particular, we do a projection that gives us an Op-plane that extends along the directions $01\ldots p$. This means that only supercharges that satisfy
\be
\epsilon_1 = \Gamma_{01\ldots p} \epsilon_2 \equiv \Gamma_{\rm Op} \epsilon_2
\ee
survive and the other 16 supercharges are projected out. Now it is meaningful to talk about a Dp-brane as the object that preserves the same 16 linear supercharges as the Op-plane, i.e. the brane that has $\Gamma_{\rm Dp} = \Gamma_{\rm Op}$. The anti-Dp-brane is the object for which the 16 linear supercharges of the background are non-linearly realized on the world volume fields.

The case of a brane at an angle relative to the Op-plane is the most interesting one. The 16 linearly realized supersymmetries of the background correspond to a combination of linear and non-linear transformations for the world volume fields on the Dp-brane. %Such a combination is certainly non-linear but it differs from the standard non-linear transformations and we will discuss this in detail in the next subsection.
Such a combination is always non-linear, as we will show in the next subsection.

Next let us discuss branes with non-vanishing world volume flux $\cF = B+F$, where $B$ is the pull-back of the background Kalb--Ramond field and $F=dA$ is the field strength on the brane. The above Dp-brane with projection condition given in equation \eqref{eq:Dp} together with the Dp-brane at an angle  $\vp$ and projection condition given in equation \eqref{eq:Dpangle} are T-dual to an D(p+1)-brane with $\cF = \tan(\vp)dx^p\w dx^{p+1}$ \cite{Johnson:2003gi}. Thus, we expect that generically branes with world volume fluxes will break the supersymmetry that is preserved by the corresponding orientifold projection. Concretely, the linearly realized supersymmetries that are preserved by a Dp-brane with world volume flux in flat space are given by (see for example section 5.3 in \cite{Koerber:2010bx} and note that in our conventions $\varepsilon^{01\ldots p}=-1$)
\be
\epsilon_1 =- \frac{1}{\sqrt{g+\cF}} \sum_{2n+\ell=p+1} \frac{1}{n!\ell!2^n} \varepsilon^{a_1 \ldots a_{2n} b_1 \ldots b_\ell} \cF_{a_1a_2} \ldots \cF_{a_{2n-1}a_{2n}} \Gamma_{b_1\ldots b_\ell} \epsilon_2 \equiv \Gamma_{\rm Dp}^\cF \epsilon_2\,,\label{epsilonprojection}
\ee
with $\Gamma_a = \partial_a y^M \Gamma_M$ being the pull-back of the Gamma matrices to the Dp-brane world volume. Since generically $\Gamma_{\rm Op} \neq \Gamma_{\rm Dp}^\cF$ the 16 linearly realized supercharges of the background are non-linearly realized on the brane with world volume flux. We will show that all such setups cannot be described using standard linear supersymmetry.

The above examples can be straightforwardly extended to compactifications of type II supergravity on spaces that partially preserve supersymmetry, like for example Calabi--Yau manifolds. In this case one has to ensure that Gauss law is satisfied and the charges carried by the branes, orientifolds and potentially background fluxes have to cancel. This leads to a large sets of possibilities: for example in compactifications to four dimensions we can use as internal space $T^6$, $T^2 \times K3$ or a CY$_3$ manifold and get $\N=8$, $\N=4$ or $\N=2$ supergravities. We can then do orientifold projections that break half or more of the supersymmetry. Added D-branes will now have world volume fields that transform linearly and/or non-linearly under the remaining supercharges.

The above branes at angles, anti-branes and branes with world volume fluxes are the main examples that we have in mind and we will discuss them in this paper. However, there are other supersymmetry breaking sources in type II like for example NS5-branes, (p,q)-branes, and KK-monopoles that can likewise be included in a supergravity action using the general idea of non-linear supersymmetry outlined below in this paper. Similarly, our discussion applies to supersymmetric M-theory or heterotic string theory compactifications with supersymmetry breaking sources in arbitrary dimensions.

\subsection{Linear vs. non-linear supersymmetry}
In the previous section we have seen that D-branes in string theory spontaneously break part or all of the supersymmetry preserved by a given background. Concretely, we have a given background with fields that transform linearly under a certain number of supercharges $\{Q^A\}$, $A=1,\ldots,N$. We add to that D-branes or other localized sources that transform linearly under a (potentially empty) subset $\{Q^a\}$, $a=1,\ldots, n<N$. What does this mean for the remaining supercharges $\{Q^i\}$, $i=n+1,\ldots,N$?

Above we have seen that for a D-brane at angles in the presence of an Op-plane, the 16 linearly realized supersymmetries of the background will combine with the non-linear supersymmetry of the world volume fields. A combination of linear and non-linear supersymmetry transformations is generically non-linear, as it can be easily shown. Consider indeed a generic situation in which a set of scalars $\phi^A$ are mapped by supersymmetry to spin-1/2 fields $\lambda^A$. The transformation of the $\lambda^A$ will be schematically of the form
\begin{equation}
\delta \lambda^A = S^A \epsilon + \slashed{\partial} \phi^A \epsilon +\dots,
\end{equation}
where $S^A$ are complex constants that are not all vanishing in any parametrization of the scalar fields. The dots stand for other model-dependent terms, which are not relevant for the present argument. We can split $S^A = \{S^a,S^i\}$, where $S^a=0$, while $S^i \neq 0$ are the fermionic shifts, which break supersymmetry spontaneously. Let us focus on the non-linear part of the supersymmetry transformations given by the shift, namely
\begin{equation}
\begin{aligned}
\delta_\text{shift} \lambda^a &=0,\\
\delta_\text{shift} \lambda^i &=S^i\epsilon.
\end{aligned}
\end{equation}
This part is rather universal and is generically present in any model with spontaneously broken supersymmetry.
We would like to show that, as long as at the set $\{S^i\}$ is not empty, then it is not possible to simultaneously set to zero all the non-linear parts  of the supersymmetry transformations, even if we allow for field redefinitions.
In other words, we show that a combination of linear and non-linear supersymmetry transformation is always non-linear. Since the set $\{\lambda^a\}$ already transforms linearly, we focus on the complementary set $\{\lambda^i\}$.
We can then define a spin-1/2 field
\begin{equation}
v =  g_{i\jb}\lambda^iS^\jb \equiv \lambda^T g \bar S,
\end{equation}
where $g_{i\jb}$ is the \K metric on the scalar manifold, such that
\begin{equation}
\delta_\text{shift} v = g_{i\jb} S^i S^\jb \epsilon \equiv (S^T g \bar S)\epsilon\,.
\end{equation}
We can redefine now all the other fields in the set  $\{\lambda^i\}$ as
\begin{equation}
\tilde \lambda^i = \lambda^i - \frac{S^i v}{S^T g \bar S},\qquad \tilde \lambda^i g_{i\jb}  S^\jb=0\,,
\end{equation}
such that
\begin{equation}
\delta_\text{shift}\tilde \lambda^i = 0.
\end{equation}
However, for this to be consistent we have to require $(S^T g \bar S) \epsilon \neq 0$ and therefore $\delta_\text{shift} v \neq 0$. Notice that the same requirement follows from the fact that the metric $g_{i\jb}$ is positive definite, which actually implies $(S^T g \bar S) = ||S||^2 >0$. This means that, even if by means of field redefinitions we can restrict the non-linearity to be present along just one direction in field space, we can never really eliminate it, since the field space metric $g_{i\jb}$ is positive definite. This argument is fully generic and can be easily adapted to any specific setup. We also stress that this argument assumes that the order parameter responsible for supersymmetry breaking contains a constant piece that cannot be removed by a field redefinition.\footnote{A simple example would be non-supersymmetric Dp-branes on tori where the order parameter would be related to the wrapping numbers of the Dp-branes.} 

This concludes a simple proof that shows that string compactifications in which supersymmetry is broken by local sources cannot be described using the language of standard linear supersymmetry or supergravity. The reason is that the combined actions for the closed string background fields and the local sources can never be invariant under linear supersymmetry, as proven above.\footnote{This statement is not in contradiction with the analysis in \cite{Cribiori:2017ngp}. Indeed, there the attention was on the non-linear and model dependent interactions in the supersymmetry transformations, rather than on the constant shift, which was always assumed to be present.} The linearly realized supersymmetries of the background action are non-linearly realized on the world volume fields of the sources and vice versa. Therefore, clearly such compactifications require that supersymmetry is non-linearly realized for some fields and for example for compactifications to four dimensions with four non-linearly realized supercharges this leads to the so-called dS supergravity theory \cite{Antoniadis:2014oya,Dudas:2015eha, Bergshoeff:2015tra, Hasegawa:2015bza} coupled to different kinds of matter and gauge fields \cite{Kallosh:2015sea, Schillo:2015ssx}.

\subsection{More details on the transition from linear to non-linear SUSY}
In the previous subsection we have proven that the supersymmetry that is spontaneously broken by branes at angles or by branes with world volume fluxes, cannot be described using standard linear supersymmetry. When supersymmetry (or any other symmetry) is spontaneously broken at a certain scale $F$, then we expect that above that scale new degrees of freedom come in and restore the linear symmetry. The necessity for such new states manifests itself in unitarity violations in for example the original Volkov-Akulov theory \cite{Volkov:1972jx, Volkov:1973ix}, as well as for example for a chiral superfield with very heavy scalar component coupled to supergravity \cite{Casalbuoni:1988sx}. So, the expectation is that non-linear supergravity theories will violate unitarity above the supersymmetry breaking scale and new degrees of freedom need to come in. One notable exception to this is inflation in non-linear supergravity, where a large Hubble scale $H$ modifies the simple argument above. In particular, the SUSY breaking scale $F\sim m_{3/2} M_{Pl}$ gets replaced with $F \sim   \sqrt{m_{3/2}^2+H^2}  M_{Pl}$ \cite{Kallosh:2000ve}, see \cite{DallAgata:2014qsj, Ferrara:2015tyn, Carrasco:2015iij, Ferrara:2016een} for a recent discussion of this point. Here we are restricting ourselves to models with negligible Hubble scale and are asking how is unitarity preserved in string theory models that give rise to non-linear supergravity theories at low energies?

Below we will study a few examples and find generically massive states with masses around the SUSY breaking scale $F$, as is required for a unitary and UV complete theory like string theory. However, generically we do not find a single set of such states but infinite towers of states whose mass scale is set by the SUSY breaking scale. Such infinite towers of states invalidate the use of the original low energy effective theory at energies above the SUSY breaking scale but in some examples one can integrate in the entire new tower of states and describe the resulting new theory. Given this observation it seems likely that any generic low-energy effective supergravity theory that arises from string compactifications in which sources (like Dp-branes) break SUSY, will only be valid below the supersymmetry breaking scale. This means that one would expect a substantial change in such a generic  low energy effective theory near the SUSY breaking scale. Here we mean by a generic theory one for which local sources are neither absent nor specially aligned such that they preserve (almost) linear supersymmetry.

The above statement might seem fairly strong with substantial implications for our universe, if it is generic in the above sense. So, let us add some words of caution. Firstly, it is not impossible that a low energy effective theory undergoes substantial changes that do not affect all of its sectors. For example, in the KKLT construction \cite{Kachru:2003aw} an anti-D3-brane at the bottom of a warped throat breaks supersymmetry spontaneously. The SUSY breaking scale is the warped down string scale, which in controlled settings should be above the warped KK scale. So, below the SUSY breaking scale new massive KK states will appear and at the SUSY breaking scale massive open string states appear. However, all these states are localized at the bottom of the throat and this might not substantially affect another sector located in the bulk or even in a different throat. Secondly, local sources minimize their energy if they preserve mutual linear supersymmetry. So, it seems possible that supersymmetry preserving sources, albeit being non generic in the above sense, are very abundant since they are dynamically favored.\footnote{It would be interesting to quantify this.} If all sources preserve linear supersymmetry and it is only broken by for example background fluxes, then we are not aware of any reason why one would expect an infinite tower of states to appear near or above the SUSY breaking scale. So, the SUSY breaking by generic branes seems somewhat different.\footnote{One can however probably argue that generic string compactifications even without branes should not have huge hierarchies nor preserve supersymmetry. So, in that sense the closed string sector is special, if it gives rise to an effective theory for which the SUSY breaking scale is well below the KK-scale and the string scale, both of which are scales at which infinite towers of states appear.}

Here we want to give a more intuitive physical explanation for why one expects an infinite tower of light states to appear near the SUSY breaking scale if SUSY is broken by local sources like D-branes. The reason is simply that the SUSY breaking scale for these objects is the string scale that sets the scale for the tower of massive string states. Even if we manage to lower this scale, which is usually taken to be fairly high in string compactifications, then we also lower the scale for the tower of light strings on the D-brane. Concretely, let us consider a D-brane (or anti-D-brane) in 10d flat space. It spontaneously breaks half of the 32 supercharges preserved by the background. The associated SUSY breaking scale is the string scale. We can project out the remaining linearly realized supercharges by combining an Op-plane with an anti-Dp-brane. The resulting theory has no linear supersymmetry and as we have proven above cannot be possibly rewritten such that one gets a theory with linear supersymmetry. Therefore, like for the original Volkov--Akulov theory, one expects that four goldstino interaction terms, that have to be present in order for the action to be invariant under non-linear supersymmetry, lead to divergent cross-sections and unitarity violations at energies well above the SUSY breaking scale. String theory cures this via the infinite tower of massive string states on the D-brane. Here notably the first level of massive open string states does not contain the right number of states to allow for linear supersymmetry restoration, so it seems likely that one needs the entire tower of massive open string states.\footnote{For example, one could consider an anti-D3-brane in flat space on an O3$^-$-plane. It is known that at the ground state we have 8 fermions, while the first excited level has 128 bosons. On the other hand, the $\mathcal{N}=4$ representation of linear supersymmetry in four dimensions has 8+8 degrees of freedom.} It would be interesting to precisely show how linear supersymmetry is restored by including the entire tower of massive open string states. We hope to come back to this in the future.

Next, we present a few examples of string compactifications with D-branes that break all the linear supersymmetry.

Our first example is the KKLT construction of dS vacua in string theory \cite{Kachru:2003aw}. Here we have a four-dimensional theory in AdS that preserves linear $\N=1$ supersymmetry. To this an anti-D3-brane is added at the bottom of a warped throat in the compactification manifold. This spontaneously breaks supersymmetry and uplifts the AdS vacuum. The uplift energy that is the SUSY breaking scale is the warped down string scale. Since the anti-D3-brane is sitting at the bottom of a warped throat the masses of the entire open string tower are warped down. Therefore, at or near the SUSY breaking scale we have the tower of massive open string states coming in, which should lead to a substantial modification of the effective low energy theory. However, here things can be more interesting. In particular for the Klebanov--Strassler throat \cite{Klebanov:2000hb}, we have an $S^3$, i.e. a non-trivial 3-cycle, at the bottom of the throat. This means that we have a warped down KK scale that in controlled setups is below the warped string scale. So, there is an infinite tower of KK states coming in before the infinite tower of massive string states on an anti-D3-brane. This was studied in great detail in \cite{Aalsma:2018pll}. The authors study the situation with many anti-D3-branes that polarize into an NS5-brane that can wrap a metastable cycle on the $S^3$ \cite{Kachru:2002gs}. It turns out that it is possible to restore linear supersymmetry by including a particular tower of KK modes at the bottom of the Klebanov--Strassler throat \cite{Aalsma:2018pll}.

Next, let us consider a D-brane that is rotated by a very small angle $\varphi \ll 1$ relative to another D-brane. The supersymmetry breaking scale is then proportional to $\varphi/\sqrt{\alpha'}$, where $1/(2\pi\alpha')$ is the string tension. By making $\varphi$ arbitrarily small, we can have an arbitrarily small SUSY breaking scale that can certainly be well below the compactification scale and the string scale set by simply $1/\sqrt{\alpha'}$. So, it might seem natural that one could describe this also in terms of a standard effective low energy supergravity theory. However, this cannot be the case as we have proven in the previous subsection. In the past such cases were often described as D-terms in standard linear 4d $\N=1$ supergravity for very small angles and we will correct this in the next section. Here we are more interested in the breakdown of such a low energy effective theory near the SUSY breaking scale. One can actually show that $\varphi/\sqrt{\alpha'}$ sets the scale of a tower of massive open string states that stretch between the two branes, see for example \cite{Berkooz:1996km, Anastasopoulos:2011hj, Anastasopoulos:2016cmg}. Therefore, there will be a full tower of excited open string modes with masses being multiples of the SUSY breaking scale. So, the effective low energy theory breaks down and it is only when one includes an infinite tower of new states that one can potentially obtain a new theory that describes this string compactification and that has linearly realized supersymmetry. In this particular setup it is of course not too difficult to guess that such a theory should be an $SU(2)$ gauge theory arising from two coincident D-branes. In that theory one can turn on a $\varphi$ dependent vev for one of the scalar fields on one of the D-branes. This rotates one of the branes and breaks the gauge group to $U(1)\times U(1)$ (see \cite{Berkooz:1996km} for some related discussions).

The above example is somewhat peculiar in the sense that we fine-tuned the intersection angle to get a very small supersymmetry breaking scale but nevertheless we ended up with an infinite tower of states. This tower is associated with strings stretching between two D-branes and therefore different from the tower of massive open string states that arise on any single D-brane and that we discussed before. Therefore, one can modify it such that one can find examples without such infinite towers of states near the SUSY breaking scale. Let us consider two intersecting Dp-branes, for example two D6-branes that intersect perpendicular on a toroidal orbifold, such that they preserve some supersymmetry. Now if we change the angle slightly away from this supersymmetric angle then this again leads to a very small SUSY breaking scale that can be well below the KK scale and the string scale. This time the intersection angle is not small and therefore there is no infinite tower of massive string states appearing near the SUSY breaking scale. This can be understood as follows. The fine tuning of the angle to be very close to the supersymmetric angle leads to a world volume theory on the slightly rotated Dp-brane that is almost invariant under linear supersymmetry. This allows the world volume fields to be almost in regular multiplets and to almost cancel each other in divergent cross-sections. So, no infinite tower of states needs to come in at the SUSY breaking scale. However, there is no way to rewrite this in terms of linearly realized supersymmetry as proven above. Therefore, there is no way to write down the exact action in this setup using standard linear supergravity. In this paper we are laying out the first steps towards a correct low energy description of such setups in the next section.

We believe that the above examples are describing a generic feature of string compactifications in which all supersymmetry is broken by D-branes, namely the appearance of a tower of states whose mass is set by the SUSY breaking scale. This means that the resulting low energy effective theories that describe generic compactifications will be substantially modified at or near the SUSY breaking scale. However, there are ways of avoiding such modifications in concrete setups by fine tuning as in the last example above.

In the next section we will discuss how to explicitly describe particular setups below the SUSY breaking scale. We will use non-linearly realized supersymmetry and see that there is no smooth limit in which we can obtain a theory with linear supersymmetry. The SUSY breaking scale will appear with inverse powers and sending it to zero will lead to singularities. Alternatively, if one redefines the field so that there are no singularities one finds that the entire action of the goldstino field is proportionally to the SUSY breaking scale and hence disappears once the SUSY breaking scale is send to zero. However, the goldstino is a fermionic field that lives on the branes and it does not disappear at higher energies. On the contrary, generically new light fields will appear for energies near or above the SUSY breaking scale. So, it does not seem possible to capture this transition from non-linear supersymmetry to linear supersymmetry in a simple low energy effective theory.

\section{Non supersymmetric branes in 4d \texorpdfstring{$\N=1$}{N=1} theories}\label{sec:nonsusybranes}
In this section we describe how to incorporate \emph{non-supersymmetric} Dp-branes in compactifications of type II string theory to four-dimensional theories with $\N=1$ supersymmetry. We have in particular flux compactifications and cosmological applications in mind \cite{Grana:2005jc, Douglas:2006es, Blumenhagen:2006ci} and we will neglect the model dependent world volume fields on the Dp-branes. However, our results are also relevant for the existing literature on intersecting D-brane model building \cite{Blumenhagen:2005mu, Blumenhagen:2006ci} and it is possible but technically challenging to include all world volume fields in any given model, as was done in \cite{Vercnocke:2016fbt, Kallosh:2016aep, Aalsma:2017ulu, GarciadelMoral:2017vnz, Cribiori:2019hod} for the anti-D3-brane in the KKLT setup \cite{Kachru:2003aw}.

\subsection{The D3-brane}
The case of the D3-brane is somehow peculiar since, in order to preserve Lorentz invariance in the non-compact space, the brane cannot be at arbitrary angles. Indeed, let us consider a generic type IIB compactification on a Calabi-Yau 3-fold. In order to break 4d $\mathcal{N}=2$ supersymmetry to $\mathcal{N}=1$ we introduce also an orientifold projection, implying
\begin{equation}
\epsilon_1 = \Gamma_{0123} \epsilon_2 \equiv \Gamma_{\rm O3} \epsilon_2.
\end{equation}
On the other hand, the presence of a D3-brane at some angle $\varphi$ in the $(x^3,x^4)$-plane requires
\begin{equation}
\epsilon_1 = \left( \cos (\varphi) \Gamma_{0123} + \sin (\varphi) \Gamma_{0124}\right)\epsilon_2 \equiv \Gamma_{\rm D3}(\varphi)\epsilon_2.
\end{equation}
Since the term proportional to $\Gamma_{0124}$ would break Lorentz invariance in the non-compact space, we have to require $\sin(\varphi)=0$. Therefore the only possible angles for a D3-brane are $\varphi=\{0,\pi\}$. In the first case, the brane preserves the same supersymmetries as the orientifold projection, while in the second case all the supersymmetries are spontaneously broken and the brane is an anti-D3-brane. Notice that this also means that, in the D3-brane case with an orientifold projection, the world volume flux $\mathcal{F}$ has to vanish for both the allowed values of $\varphi$.

\subsection{Higher-dimensional supersymmetric branes in type II}
For a compactification to four dimensions we can include Dp-branes with $p>3$ that extends along the non-compact spacetime directions and wrap an internal $p-3$-cycle $\Sigma$. The action of such a Dp-brane is given in string frame by
\ba
S_{\rm Dp} &=& S_{\rm DBI,p}+S_{\rm CS,p} \cr
&=& T_{\rm Dp} \lp \int_{M^{3,1}\times \Sigma} d^4x\, d^{p-3}y\,  e^{-\phi|_\Sigma} \sqrt{\det \lp -g|_\Sigma + \cF|_\Sigma\rp} - \int_{M^{3,1}\times \Sigma} C|_\Sigma \w e^{\cF|_\Sigma} \rp\,,\quad \label{eq:Dpaction}
\ea
where $T_{\rm Dp}$ is the brane tension, $\phi$ the dilaton, $g$ the metric and $C$ is sum over the RR-fields. We also denoted the pull-back onto the world volume of the Dp-brane by $|_\Sigma$. For the ease of the notation we will not spell out the pull-back anymore and we restrict in this paper to compactifications that preserve $\N=2$ supersymmetry in four dimensions, broken to $\N=1$ after doing an orientifold projection. In the case with vanishing fluxes the internal space is therefore a CY$_3$ manifold and otherwise it is a more general SU(3)-structure manifold. These spaces (in the strict limit) have no non-trivial 1- and 5-cycles and are equipped with a \K (1,1)-form and a holomorphic (3,0)-form $\Omega$.\footnote{Generic SU(3)-structure manifolds are not \K and the \K form $J$ and holomorphic three-form $\Omega$ are not closed so we are slightly abusing the language here. The existence of the real 2-form $J$ and the complex 3-form $\Omega$ are a defining property of SU(3)-structure manifolds as explained for example in section 3.2 of \cite{Grana:2005jc}. In the special case of Calabi-Yau manifolds they reduce to the familiar \K and holomorphic 3-form.} Supersymmetric Dp-branes in such compactifications have been reviewed in great detail in \cite{Blumenhagen:2006ci}. It turns out that one can express the DBI part of the action for supersymmetric branes in terms of the \K form $J$, the holomorphic 3-form $\Omega$ and the Kalb-Ramond 2-form $B$ as follows\footnote{We will not discuss the CS-part of the action much further here. It encodes the charge of the Dp-brane and enters in the tadpole cancelation condition that ensures that the total charge in the internal compact space vanishes, if one combines the contribution from fluxes, D-branes and orientifold planes.}
\ba
S_{\rm DBI,5} &=& T_{\rm D5}  \int_{M^{3,1}\times \Sigma_2} d^4x\, e^{-\phi} J\,,\cr
S_{\rm DBI,6} &=& T_{\rm D6}  \int_{M^{3,1}\times \Sigma_3} d^4x\, e^{-\phi} \Re(\Omega) \,,\cr
S_{\rm DBI,7} &=& T_{\rm D7}  \int_{M^{3,1}\times \Sigma_4} d^4x\, e^{-\phi} \frac12 \lp J\w J-B\w B\rp\,,\cr
S_{\rm DBI,9} &=& T_{\rm D9}  \int_{M^{3,1}\times \Sigma_6} d^4x\, e^{-\phi} \lp \frac16 J\w J\w J-\frac12 J\w B\w B \rp\,,\label{eq:calibration}
\ea
where we used string frame and have set the world volume gauge flux $F=0$, as well as all other world volume fields on the Dp-branes.

The above calibration conditions have already been generalized to supersymmetric branes with non-zero world volume flux $F$, see for example \cite{Martucci:2005ht, Martucci:2006ij, Martucci:2011dn}. Such supersymmetric Dp-branes satisfy generalized calibration conditions and it is possible to use them to rewrite the DBI-action in equation \eqref{eq:Dpaction} in terms of expressions that generalize the expressions in \eqref{eq:calibration}. In particular, for any Dp-brane wrapping a ($p-3$)-cycle $\Sigma$ one finds that the DBI action contributes the following terms to the scalar potential \cite{Martucci:2005ht, Martucci:2006ij}
\be
V_{\rm DBI} = \int_\Sigma e^{4A-\phi}\  \Re \hat \Psi_1 \w e^{\cF}\,.
\ee
Here $e^{4A}$ denotes the warp factor and $\hat \Psi_1$ is a pure spinor (which is a polyform). The definition of the pure spinor is different in type IIA and type IIB so that the above expression reduces to equation \eqref{eq:calibration} for vanishing world volume flux and no warping.

In this paper we are interested in non-supersymmetric Dp-branes and will describe the general procedure of how to incorporate these into an effective low energy supergravity action in the next subsection. Then we will discuss several examples where we apply this procedure to obtain proper dS supergravities that incorporate the universal contributions from these non-supersymmetric Dp-branes.

The particular case of so-called pseudo-calibrated anti-Dp-branes in flux compactifications, first introduced on $SU(3)\times SU(3)$ structure manifold in \cite{Lust:2008zd}, was recently studied in \cite{Kallosh:2018nrk}. There the contribution to the action from these anti-Dp-branes is essentially the same as above in equation \eqref{eq:calibration}. The only difference is an overall minus sign in the CS-term since the anti-Dp-branes have the opposite orientation, so their volume, as measured by the DBI-action is the same as for Dp-branes. Crucially however, for the supersymmetric Dp-branes the cancelation of the Dp-brane charge ensures that the above DBI-action nicely fits into the standard linear supergravity formalism, while for anti-Dp-branes this is not the case and one finds new terms of the form given in equation \eqref{eq:calibration} that can only be incorporated into the scalar potential when using non-linear supergravity, see \cite{Kallosh:2018nrk} for details. Here we are interested in more general non-supersymmetric Dp-branes and we will generalize the results of \cite{Kallosh:2018nrk}.

\subsection{The new supergravity action}
In this subsection we show how one can generically incorporate the new contributions to the scalar potential that arise from the non-supersymmetric Dp-branes. The world volume fields that arise on a particular Dp-brane are model dependent and cannot be spelled out in full generality.\footnote{They have been studied for supersymmetric branes in for example \cite{Jockers:2004yj, Grimm:2011dx, Kerstan:2011dy, Carta:2016ynn, Escobar:2018tiu} and even there they are not fully understood in all setups.} However, all the non-supersymmetric branes that we study have in common that they break supersymmetry spontaneously and lead to a non-linear realization. This means that the world volume fields are not appearing in standard multiplets anymore. In particular, there is one special so called nilpotent chiral multiplet $S$, that satisfies $S^2=0$. This multiplet has only one fermionic degree of freedom $\psi$ because the nilpotency condition fixes the scalar in the multiplet $S=\phi + \sqrt{2} \psi\, \theta + F \,\theta^2$ in terms of the fermion as $\phi = \psi^2/(2F)$. Here $\theta$ represents the superspace coordinates and $F$ is an auxiliary field. Note, that this solution for $\phi$ necessarily requires that $F\neq 0$ and we cannot take the limit of sending $F$ to zero.\footnote{If one rescales the fermions such that $\psi\rightarrow\sqrt{F}\, \psi'$ then the entire action for the rescaled fermion $\psi'$ vanishes in the limit of $F \rightarrow 0$. However, $\psi'$ is a world volume field and cannot simply disappear, if we for example rotate the D-brane.} So, we see explicitly in this example that one cannot really take the limit that restores linear supersymmetry.

Nilpotent superfields were first used in \cite{Rocek:1978nb} in order to linearize the Volkov-Akulov (VA) model \cite{Volkov:1973ix}. The description of the VA-model in terms of nilpotent superfields serves as an illustration of how these fields realize non-linear supersymmetry. The action of the VA-model reads
\be
S_{VA} = -M^4 \int E^0 \wedge E^1 \wedge E^2 \wedge E^3 \qquad \text{with} \qquad E^\mu = dx^\mu + \bar{\lambda} \gamma^\mu d\lambda
\ee
and is invariant under the non-linear symmetry transformation
\be
\delta_\epsilon \lambda = \epsilon +\left( \bar{\lambda} \gamma^\mu \epsilon \right) \partial_\mu \lambda\,.
\ee
Using the nilpotent field $S$ this action can be written as\footnote{Note that the fermion $\lambda$ in the VA-action and the fermion $\psi$ in the nilpotent chiral superfield are not the same but are rather related via a field redefinition, see \cite{Kuzenko:2011tj, Bergshoeff:2015jxa}.}
\be
S=\int d^4x \int d^2 \theta \int d^2\bar{\theta} S\bar{S} + M^2 \left( \int d^4x \int d^2 \theta S+h.c. \right)\,,
\ee
where we now also have to consider the superspace integral over the $\theta$ coordinates. One can show that the two actions are equivalent, if one imposes the nilpotent condition $S^2=0$. In this description $\psi$ is the goldstino and the only degree of freedom. Note that this description of the VA action using a nilpotent chiral field is not a unique choice and one can also use a constrained vector multiplet \cite{Lindstrom:1979kq,Samuel:1982uh}. In addition to the nilpotent field we consider here there are many more constrained supermultiplets that can be used to describe non-linearly realized supersymmetry \cite{Samuel:1982uh, Casalbuoni:1988xh, Komargodski:2009rz, Kuzenko:2011ti, Kuzenko:2011tj, Vercnocke:2016fbt, DallAgata:2016syy, Bandos:2016xyu, Buchbinder:2017vnb}. A description on how to use these fields in supergravity can be found in \cite{DallAgata:2015zxp}. Constrained multiplets have been used to great success in order to describe the complete action of the anti-D3-brane in the KKLT background \cite{GarciadelMoral:2017vnz,Cribiori:2019hod} and general anti-Dp-branes in \cite{Kallosh:2018nrk}. In \cite{Cribiori:2019bfx} the constrained multiplet formalism has been used in order to facilitate the first uplift in type IIA, using anti-D6-branes.

After this short review of constrained multiplets, let us return to describing the low energy contribution of supersymmetry breaking D-branes in four-dimensional $\N=1$ supergravities. If the background preserves linear $\N=1$ supersymmetry, then the added non-supersymmetric branes would be the sole source of supersymmetry breaking. That means that non-supersymmetric branes would have to provide the Goldstino, i.e. one of the world volume fermions or a linear combination thereof is the Goldstino. We can incorporate this Goldstino into our general action via the nilpotent chiral multiplet $S$. This allows us to incorporate the new contribution from the non-supersymmetric branes into the bosonic supergravity action as follows: 

We start with any four dimensional theory with linear $\N=1$ supersymmetry with a K\"ahler potential $K_{\rm before}$, superpotential $W_{\rm before}$, as well as potentially gauge kinetic functions and D-terms. These are obtained from an explicit string theory compactification and depend on chiral multiplets $\Phi^a$. We now add non-supersymmetric Dp-branes and we want to incorporate their backreaction on the $\Phi^a$ fields. This backreaction can be obtained by explicitly reducing the non-supersymmetric Dp-brane actions in a particular compactification, which gives rise to a new scalar potential term $V_{\rm new}(\Phi^a,\bar \Phi^{\bar a})$. We now define the full K\"{a}hler potential and superpotential as\footnote{Inflationary models based on similar modifications of the K\"ahler potential were first studied in \cite{McDonough:2016der}.}
\begin{eqnarray}
K&=&K_{\rm before} + \rme^{K_{\rm before}} \frac{S \bar S}{V_{\rm new}}\,,\nonumber\\
W&=&W_{\rm before} + S\,. \label{eq:KWmodified}
\end{eqnarray}
This leads to the following scalar potential with a generic new contribution from the non-supersymmetric branes
\be
V=V_{\rm F}+V_{\rm D}=\rme^K\left.\lp K^{I\bar J} D_I W \overline{D_J W}-3|W|^2 \rp\right|_{S=0} +V_{\rm D}=  V_{\rm before} + V_{\rm new}\,.\label{eq:Vmodified}
\ee
Here $I, J$ run over all fields including the constrained nilpotent chiral multiplet $S$. Since $S$ only contains the Goldstino we have to set it to zero in the end to obtain the bosonic scalar potential. We thus have $D_S W|_{S=0}=1$ and
\begin{equation}
K^{S\bar S}\big|_{S=0}=\left(K_{S\bar S}\big|_{S=0}\right)^{-1}= \rme^{-K_{\rm before}} V_{\rm new}\,.
 \label{KSbarS}
\end{equation}
%and this removes all derivatives of $V_{\rm new}$ in the full scalar potential.

The above actually works independently of whether the background has supersymmetry preserving solutions or not and it will give us always the correct result. It is however not always immediately obvious that $V_{\rm new}$ is a real function of the $\Phi^a$. We will argue that this is always the case and this might also be intuitively clear because $V_{\rm new}$ is a real function of the closed string degrees of freedom that have been package into the $\Phi^a$. So, we conclude that we can always incorporate the new contributions from non-supersymmetric branes by using a nilpotent chiral multiplet. Generically, the background fields will also break supersymmetry, in particular once we include the backreaction of the non-supersymmetric branes. This means that the Goldstino is not simply the fermion contained in $S$ but rather a linear combination like $F_S \lambda_S + F_\alpha \lambda_\alpha$, where the index $\alpha$ runs over the chiral multiplets that arise from the closed string sector.

Our prescription above seems rather simple and ad hoc. However, it was shown, and we will review this below, that it gives the correct description in all examples. This might seem trivial since $V_{\rm new}$ can be any real function of the other moduli. However, the changes to $K$ and $W$ become highly non-trivial, if one wants to include all world volume fields on a Dp-brane and reproduce all couplings between the bosonic and fermionic world volume fields and the closed string background fields. This has only been worked out in full detail for an anti-D3-brane in the KKLT setup \cite{GarciadelMoral:2017vnz, Cribiori:2019hod}. The rather general elaborate answer in this case, see section 5 of \cite{Cribiori:2019hod}, reduces exactly to the expression above, if we set all world volume fields but the Goldstino to zero.\footnote{It would be interesting and probably not too difficult to include the scalar degrees of freedom from the world volume fields on the non-supersymmetric branes. These appear already in $V_{\rm new}$. One would have to add more constrained multiplets to the original theory and include them in $K$ such that one obtains the correct kinetic terms for the world volume scalars.} The simplicity and elegance of the equations \eqref{eq:KWmodified} and \eqref{eq:Vmodified} can also be understood from the fact that they use the universal feature of all SUSY breaking Dp-branes, namely the presence of a Goldstino contained in a nilpotent multiplet $S$, and package everything else, like information about the compactification background, the dimension of the brane and so on, into the unspecified function $V_{\rm new}$. Below we will see what $V_{\rm new}$ is in several concrete examples and show that the correct scalar potential can be obtained as described above.

\subsection{Explicit examples of genuine non-supersymmetric branes}

\subsubsection{The anti-D3-brane}
The anti-D3-brane is the simplest and most studied case of supersymmetry breaking by Dp-branes, since the anti-D3-brane plays an important role in the KKLT scenario \cite{Kachru:2003aw}, where it was used to uplift a supersymmetric AdS vacuum to a dS vacuum. In the original paper the anti-D3-brane contribution was taken to be
\be
V_{\rm \overline{D3}} = \frac{\mu^4}{(-\rmi (T-\bar T))^3}\,,\label{VD33}
\ee
which is the correct result for a generic anti-D3-brane in a compactification with a single K\"ahler modulus $T$. Here $\mu$ is related to the tension of the anti-D3-brane. In KKLMMT \cite{Kachru:2003sx} it was shown that the correct result for an anti-D3-brane in a warped throat is
\be
V^{w}_{\rm \overline{D3}} = \frac{\mu^4}{(-\rmi (T-\bar T))^2}\,.\label{VDw2}
\ee
Both of these results can be reproduced as described in the previous subsection by setting either $V_{\rm new}=V_{\rm \overline{D3}}$ or $V_{\rm new}=V^{w}_{\rm \overline{D3}}$ as first noticed in \cite{Ferrara:2014kva}, leading respectively to
\begin{align}
  K=& -3 \ln \ls-\rmi (T-\bar T)\rs+ \mu ^{-4}S\bar S \,,\nonumber\\
  K=& -3 \ln \ls-\rmi (T-\bar T)\rs+ \mu ^{-4}\frac{S\bar S}{-\rmi (T-\bar T)}= -3 \ln\ls -\rmi (T-\bar T)-\frac{1}{3\mu ^4}S\bar S\rs\,,
 \label{KantiD3}
\end{align}
where we used $S^2=0$ in the last equality.

The connection between the above description and an actual anti-D3-brane was first established in \cite{Kallosh:2014wsa, Bergshoeff:2015jxa} and this has ultimately led to an explicit reduction of the full anti-D3-brane action, coupled to all background fields in \cite{GarciadelMoral:2017vnz, Cribiori:2019hod}. This gives us confidence that with a sufficient amount of effort one can likewise extend any other explicit example to a full model that contains all world volume fields on the Dp-brane. Furthermore, it should be clear from the above example that our procedure is guaranteed to give the correct answer in the limit where we set all world volume fields on the Dp-brane to zero. In that limit one just has to equate the model-dependent new contribution with our $V_{\rm new}$.

\subsubsection{Intersecting D6-branes}
The first new example to which we apply our above procedure are intersecting D6-branes in type IIA string compactifications to four dimensions. Such intersecting branes have led to standard like models, as for example reviewed in \cite{Blumenhagen:2005mu}, and they have been included in flux compactifications, as for example reviewed in \cite{Blumenhagen:2006ci}. Here we are particularly interested in brane setups that do not preserve the linear $\N=1$ supersymmetry of the background. Such setups have been argued to lead to D-term breaking potentially going back more than twenty years ago and we would like to clarify here the following:
\begin{enumerate}
\item The papers we have been looking at are studying such setups using 10d supergravity or even string theory and are using the correct DBI and worldsheet (WS) action for the D6-branes. So, the obtained results are as far as we checked all correct.
\item The repackaging or interpretation of the four-dimensional scalar potential arising from non-supersymmetric D-branes as D-terms is not correct.
\end{enumerate}
The reason, as argued above, is that we cannot use the language of linear supersymmetry since the world volume fields on the D-branes transform non-linearly and this cannot be changed or undone by any field redefinition.

A beautiful paper that discusses the contribution of D6-branes in type IIA flux compactifications is \cite{Villadoro:2006ia}. The authors focus in particular on D6-branes wrapping an arbitrary 3-cycle $\Sigma_3$ on $T^6/\mathbb{Z}_2\times\mathbb{Z}_2$. Defining\footnote{We are following the conventions of for example \cite{Blumenhagen:2006ci, Kallosh:2018nrk}. These differ from the conventions used by Villadoro and Zwirner (VZ) in \cite{Villadoro:2006ia}, so that our $\Omega$ satisfies $\Omega=\rmi \Omega^{\rm VZ}$, and our $\Omega _\Sigma $ is $\tilde \Omega _\pi $ in \cite{Villadoro:2006ia}.}
\be
\Omega_\Sigma = \int_{\Sigma_3} e^{-\phi}\, \Omega\,,
\label{defOmegaSigma}
\ee
they find that the scalar potential contribution from the DBI action takes the form\footnote{In \cite{Villadoro:2006ia}, $N$ D6 branes were considered, which just changes all $T_{D6}$ below to $NT_{D6}$.}
\be
V_{\rm DBI} = T_{D6}\,\hat{s}^{-2} \sqrt{\lp \Re \Omega_\Sigma\rp^2 +\lp \Im\Omega_\Sigma\rp^2}\,,\label{eq:D6DBI}
\ee
where $\hat{s}=e^{-2\phi} vol_6$ is a real combination of the dilaton and geometric moduli. For supersymmetric D6-branes $\Im\Omega_\Sigma =\int_{\Sigma_3}e^{-\phi}\, \Im \Omega=0$, so that $V_{\rm DBI} = T_{D6}\,\hat{s}^{-2} \Re \Omega_\Sigma$, where we used that $\Re \Omega_\Sigma>0$ for a supersymmetric brane. This can be derived from the D6-brane DBI action in string frame in equation \eqref{eq:calibration} by going to 4d Einstein frame. In particular, the 10d supergravity action contains the term
\begin{equation}
  \int d^{10}x \sqrt{-g^s_{(10)}} e^{-2\phi} R_{(10)} = \int d^4x \sqrt{-g^s_{(4)}} e^{-2\phi} vol_6 R_{(4)} + \ldots\,.
 \label{4dRstringframe}
\end{equation}
We now only rescale the 4d metric $g^s_{\mu\nu} \to \hat{s}^{-1} g^E_{\mu\nu}$, with again $\hat{s}=e^{-2\phi} vol_6$. This takes us to the 4d Einstein frame. For the Dp-brane actions given in equation \eqref{eq:calibration} this means that they all pick up an extra factor of $\hat{s}^{-2}=e^{4\phi}/(vol_6)^2$ from rescaling the 4d part of the metric only, which leads to the above result for the scalar potential.

Following \cite{Blumenhagen:2002wn} the authors of \cite{Villadoro:2006ia} decompose the above into a putative F-term and D-term contribution
\ba
V_{\rm DBI} &=& V_F + V_D\,,\cr
V_F &=& T_{D6}\,\hat{s}^{-2} \,\Re \Omega_\Sigma\,,\cr
V_D &=& T_{D6}\,\hat{s}^{-2} \lp \sqrt{\lp \Re \Omega_\Sigma\rp^2 +\lp \Im\Omega_\Sigma\rp^2} -\Re \Omega_\Sigma \rp\,.\label{eq:VD1}
\ea
This decomposition is such that $V_D=0$ for supersymmetric D6-branes ($\Im \Omega_\Sigma=0$), in accordance with the fact that D-terms cannot uplift the vacuum energy without breaking supersymmetry.
Such an idea, that we can write the SUSY breaking contributions from non-supersymmetric D-branes as a D-term in four-dimensional $\N=1$ supergravity, can be traced back twenty years to papers like \cite{Kachru:1999vj, Cvetic:2001nr}. However, given our improved understanding of non-linear supergravity and its connection to D-branes, we can correct this and give a proper description of the low energy four-dimensional effective theory for such brane setups.

As was noticed in \cite{Villadoro:2006ia}, one can attempt to describe the above D-term scalar potential in terms of a gauge kinetic function $f$ and a D-term $D$. The U(1) symmetry is the axion shift symmetry as deduced from the vector coupling in the Chern-Simons term of the D6 brane, normalized with $\mu _6=T_{D6}$. It turns out that its moment map is related to the imaginary part of (\ref{defOmegaSigma}):
\begin{equation}
  T_{D6}\Im\Omega_\Sigma= \hat{s}{\cal P}\,.
 \label{relImOmtoP}
\end{equation}
The DBI action leads to a kinetic term for the vector that would identify in the ${\cal N}=1$ formulation
\begin{equation}
  \Re f = T_{D6}\sqrt{\lp \Re \Omega_\Sigma\rp^2 +\lp \Im \Omega_\Sigma\rp^2}= T_{D6}\Re \Omega_\Sigma+ {\cal O}\left(\frac{\Im \Omega_\Sigma}{\Re\Omega_\Sigma}\right)\,.
 \label{Reffull}
\end{equation}
By omitting the second part, they identify a holomorphic function of the moduli, whose real part agrees with the leading part in (\ref{Reffull}) and whose imaginary part agrees with the $F\wedge F$ terms in the Chern--Simons term. Using that holomorphic $f$ allows them to rewrite the last line of (\ref{eq:VD1}) as (adapted from equation (2.23) in \cite{Villadoro:2006ia})
\begin{align}
V_D  =& T_{D6}\,\hat{s}^{-2}\frac{\left(\Im \Omega_\Sigma\right)^2}{\Re\Omega_\Sigma}\frac{1}{1+\sqrt{1+\lp \frac{\Im \Omega_\Sigma}{\Re \Omega_\Sigma}\rp^2}}\nonumber\\
=& \frac1{2 \Re f}\, {\cal P}^2\, \frac{2}{1+\sqrt{1+\lp \frac{\Im \Omega_\Sigma}{\Re\Omega_\Sigma}\rp^2}} \,.\label{eq:VD2}
\end{align}

The authors of \cite{Villadoro:2006ia} note that this is only ``compatible with the standard formula of $\N=1$ supergravity'', if $|\Im \Omega_\Sigma|/|\Re\Omega_\Sigma| \ll 1$. However, the mathematically rigorous statement is that the expression above takes the form of a standard D-term, iff $\Im \Omega_\Sigma=0$.  The latter is true for supersymmetric D6-branes in which case the D-term contribution in equation \eqref{eq:VD1} vanishes.\footnote{Interestingly, the authors of \cite{Villadoro:2006ia} point out that the discrepancy should be cured by including higher derivatives interactions into the DBI action. In this respect, non-linear supersymmetry is providing precisely the required higher-derivative interactions to make the whole description consistent. This supports the argument in \cite{Villadoro:2006ia}.} Another interesting case are anti-D6-branes, which are wrapping supersymmetric cycles with the opposite orientation and therefore also have $\Im\Omega_\Sigma=0$. Their contribution to the scalar potential for general flux compactifications was derived, using the language of non-linear supergravity, in \cite{Kallosh:2018nrk}. Here we generalize the results of \cite{Kallosh:2018nrk} to arbitrary D6-branes by noting that the above DBI action in equation \eqref{eq:D6DBI} can be added to any existing scalar potential by simply modifying the K\"ahler and superpotential as
\begin{eqnarray}
K&=&K_{\rm before} + e^{K_{\rm before}} \frac{S \bar S}{ T_{D6}\,\hat{s}^{-2} \sqrt{\lp\Re\Omega_\Sigma\rp^2 +\lp \Im\Omega_\Sigma\rp^2} }\,,\nonumber\\
W&=&W_{\rm before} + S\,. \label{eq:KWmodified2}
\end{eqnarray}
Here $S$ is a nilpotent chiral multiplet satisfying $S^2=0$. The only dynamical field in $S$ is the goldstino that necessarily has to live on the D6-brane world volume since the D6-brane is breaking supersymmetry. Other world volume fields like the gauge field or scalar fields living on the D6-brane are model dependent. One should be able to explicitly include them for any given setup using additional constrained $\N=1$ multiplets. Here we are not pursuing this road but would like to quickly discuss the generalization to multiple D6-branes. In this case one expects the following contribution to the scalar potential, generalizing the expression above in equation \eqref{eq:D6DBI},
\be
V_{\rm DBI} = T_{D6} \,\hat{s}^{-2} \sum_i \sqrt{\lp \Re\Omega_{\Sigma^{i}}\rp^2 +\lp \Im\Omega_{\Sigma^{i}}\rp^2}\,,\label{eq:D6sDBI}
\ee
where $i$ denotes the sum over the set of non-supersymmetric D6-branes. This can likewise be obtained by modifying the K\"ahler and superpotential as
\begin{eqnarray}
K&=&K_{\rm before} + e^{K_{\rm before}} \frac{S \bar S}{ T_{D6}\,\hat{s}^{-2} \sum_i \sqrt{\lp \Re\Omega_{\Sigma^{i}}\rp^2 +\lp\Im\Omega_{\Sigma^{i}}\rp^2}} \,,\\
W&=&W_{\rm before} + S\,.
\end{eqnarray}
Now the goldstino contained in the nilpotent chiral superfield $S$ is a linear combination of world volume fermions from all the D6-branes that break the four-dimensional $\N=1$ supersymmetry preserved by the background.

There is a point that we have omitted so far, namely we have to ensure that $V_{\rm new}$, which is $V_{\rm DBI}$ in equations \eqref{eq:D6DBI} or \eqref{eq:D6sDBI}, is actually a function of the closed string moduli. That this is the case in the concrete model studied in \cite{Villadoro:2006ia} can be seen for example from their equation (2.9). We can also show that this is generically the case for non-supersymmetric D6-branes: Using the conventions of appendix A.1 in \cite{Kallosh:2018nrk} and identifying their $e^{-\phi_4}$ with our $\sqrt{\hat s}$, the complex structure moduli in type IIA are given by $\mathcal{Z}^N = \int_{\Sigma^{N}} \lp \frac12 C_3 + \rmi \sqrt{\hat s} \Re \Omega\rp$, where the $\Sigma^N$ are a basis of the orientifold odd 3-homology. This means that $\Im\mathcal{Z}^N=\int_{\Sigma^{N}} \sqrt{\hat s} \Re \Omega$. It is easy to convince oneself that the overall volume and the dilaton are always real functions of the moduli, so $\hat s=e^{-2\phi} vol_6$ is also a real combination of the closed string moduli. However, the complex structure moduli $\mathcal{Z}^N$ do not involve the imaginary part of $\Omega$. So, how can we write $\Im\Omega_\Sigma$ and $\Im\Omega_{\Sigma^{i}}$ in equations \eqref{eq:D6DBI} or \eqref{eq:D6sDBI} as a real function of the complex structure moduli? The answer lies in the involution in the O6-orientifold projection. It actually acts on $\Omega$ as $\sigma: \Omega \to \overline{\Omega}$. Due to the complex conjugation in the action of $\sigma$ we can write $\Im\Omega_{\Sigma^{N}}$ as a real function of $\Re\Omega_{\Sigma^{N}}$, as explained for example in subsection 3.1 of \cite{Grimm:2004ua}, where we set their phase $\theta=0$. So, we conclude that the contribution $V_{\rm new}$ arising from the DBI actions of non-supersymmetric D6-branes is indeed a real function of the closed string moduli.

\subsection{The general answer for any Dp-brane}
Here we show how our result applies to a generic Dp-brane that spontaneously breaks the four-dimensional linear supersymmetry preserved by flux compactifications of type IIA or type IIB string theory. In particular, we will use results from \cite{Martucci:2005ht, Martucci:2006ij} and follow their notation. The Ansatz for the ten dimensional metric is
\be
ds^2 = e^{2A(y)} dx^\mu dx_\mu + g_{mn}(y) dy^m dy^n\,.
\ee
The internal space is assumed to have an $SU(3) \times SU(3)$ structure, which allows one to define two pure spinors $\hat \Psi_1$ and $\hat \Psi_2$ that differ depending on whether we are studying type IIA or type IIB (see \cite{Martucci:2005ht} for details). When considering a Dp-brane extending along the four non-compact directions and wrapping a ($p-3$)-cycle $\Sigma$ one can define
\ba
{\cal W}_m d\sigma^1\w\ldots\w d\sigma^{p-3} &=& \frac{(-1)^p}{2}\left[e^{3A-\phi} (i_m+g_{mk}\, dy^k \w) \hat \Psi_2 \right]_\Sigma \w e^\cF\Big|_{p-3}\,,\cr
{\cal D} d\sigma^1\w\ldots\w d\sigma^{p-3} &=& \left[e^{4A-\phi} \Im \hat\Psi_1\right]_\Sigma \w e^\cF\Big|_{p-3}\,,\cr
\Theta d\sigma^1\w\ldots\w d\sigma^{p-3} &=&\left[e^{4A-\phi} \Re \hat\Psi_1\right]_\Sigma \w e^\cF\Big|_{p-3}\,,\label{eq:WDT}
\ea
where the subscripts $\Sigma$ denotes the pullback to the world volume and the subscript ($p-3$) means that we only keep the ($p-3$)-form part of the expression. The DBI part of the brane action can then be rewritten as (see equations (3.1) and (3.2) in \cite{Martucci:2006ij})
\ba
S_{\rm DBI,p} &=& T_{\rm Dp} \int_{M^{3,1}\times \Sigma} d^4x\, d^{p-3}y\,  e^{-\phi|_\Sigma} \sqrt{\det \lp -g|_\Sigma + \cF|_\Sigma\rp} \cr
&=&  T_{\rm Dp} \int_{M^{3,1}\times \Sigma} d^4x\, d^{p-3}\sigma \sqrt{\Theta^2+e^{4A} {\cal D}^2 + 2 e^{2A} g^{mn} {\cal W}_m {\cal W}_n}\,.\label{eq:genericDBI}
\ea
The above expression reduces to the previous result for D6-branes given in equation \eqref{eq:D6DBI}, if ${\cal W}_m=0$ and after going to 4d Einstein frame as discussed below equation \eqref{eq:D6DBI}. Similarly, it reproduces the result for non-supersymmetric D7-branes in type IIB O3/O7 flux compactifications as given for example in equations (5) and (6)  of \cite{Palti:2014kza}, if we set ${\cal W}_m=0$. So, let us try to understand how our approach is compatible with the general result for any Dp-brane in an $SU(3) \times SU(3)$ manifold, i.e. why is ${\cal W}_m=0$ for us and why is $\Theta^2+e^{4A} {\cal D}^2$ a function of the closed string moduli?

We have seen above that we can in principle get pretty much any new contribution to the scalar potential in equation \eqref{eq:Vmodified} by adding the nilpotent field $S$ and changing the K\"ahler and superpotential as in equation \eqref{eq:KWmodified}. However, the field $S$ contains only the goldstino field out of all world volume fields on the Dp-brane. So, our approach needs to be extended if we want to include all open string fields. This seems like a daunting task, if one explicitly works out the component field action for a particular Dp-brane in a particular flux compactification including all the fermionic terms, since the latter are not fixed by non-linear supersymmetry. However, it might be possible to find a more elegant way that as the expression above captures all world volume fields. We are not trying this here but again restrict to the case where we set all world volume fields except the goldstino to zero. Then we use the fact that ${\cal W}_m$ is the derivative of a holomorphic superpotential ${\cal W}$ that encodes the open string fields \cite{Martucci:2006ij}. This means ${\cal W}_m=0$ for us, since we have set the world volume fields to zero. However, this is not sufficient for our method in equation \eqref{eq:KWmodified} to work. We also need to show that $V_{\rm new}$ is a (real) function of the closed string moduli since we included it in the K\"ahler potential. For the generic case above we see that this is the case as follows: The complex closed string moduli, universally denoted by $\Phi^a$, have in our conventions as real part combinations of the RR axions and as imaginary part NSNS fields. These imaginary parts, $\Im\Phi^a$, are directly obtained from expanding the polyform $\Re \hat\Psi_1$ in cohomology. Similarly to the type IIA example above, we have in full generality that $\Im \hat\Psi_1$ and $\Re \hat\Psi_1$ are directly related to each other as explained at the end of Section 5.1 and in footnote 21 of \cite{Grana:2005jc}, see also \cite{Grana:2005ny, Benmachiche:2006df}. This allows us to write schematically
\ba
S_{\rm DBI,p} &=&  T_{\rm Dp} \int_{M^{3,1}\times \Sigma} d^4x\, d^{p-3}\sigma \sqrt{\Theta^2+e^{4A} {\cal D}^2 }\cr
&=& T_{\rm Dp} \int_{M^{3,1}\times \Sigma} d^4x\, d^{p-3}\sigma \sqrt{f\lp\Im\Phi^a\rp +g\lp\Im \Phi^a\rp}\,,
\ea
where $f\lp\Im\Phi^a\rp$ and $g\lp\Im\Phi^a\rp$ are real functions and for supersymmetric Dp-branes one would have $g\lp{\rm Im}\Phi^a\rp=e^{4A} {\cal D}^2=0$.
So, this shows that our approach seems to be very broadly applicable and to extend beyond the most well studied classes of Calabi--Yau compactification or $SU(3)$-structure manifolds to the most general case of $SU(3)\times SU(3)$ structure compactifications of type IIA and type IIB string theory.

\subsection{A further extension}
The above general expression given in equation \eqref{eq:genericDBI} has a non-vanishing contribution in the supersymmetric limit where ${\cal D}={\cal W}_m=0$. This means that it applies to branes that can in principle be supersymmetric in the particular background. Examples are D3- and D7-branes in type IIB compactification with an O3/O7 orientifold projection or D5- and D9-branes in type IIB with an O5/O9 orientifold projection. We have described above how to incorporate such branes in a four dimensional non-linear $\N=1$ supergravity action, even if they do not wrap supersymmetric cycles and ${\cal D} \neq 0$. However, our approach of including non-supersymmetric Dp-branes in a four dimensional non-linear $\N=1$ supergravity action applies more broadly. 

Let us look at one more example to understand this: If we study type IIB compactifications with an O3/O7 orientifold projection, then any D5-brane will break supersymmetry. Nevertheless, we can include the backreaction of such non-supersymmetric D5-branes in this setup by using our formalism. The four dimensional $\N=1$ closed string moduli are 
\ba
\tau &=& C_0 +\rmi e^{-\phi}\,,\cr
G_a &=&\int_{\Sigma^a_{(2)}}C_2 +\tau B\,,\cr
T^a &=& \int_{\Sigma_a^{(4)}}\ls C_4 +C_2 \w B_2 +\frac12 C_0 B\w B -\frac{\rmi e^{-\phi}}{2} \lp J\w J - B \w B\rp \rs\,.
\ea
A D5-brane could wrap any 2-cycle and have for example, see equation \eqref{eq:calibration} above,
\be
S_{\rm DBI,5} = T_{\rm D5}  \int_{M^{3,1}\times \Sigma_2} d^4x\, e^{-\phi} J\,.
\ee
Such a D5-brane would be supersymmetric, if we had done an O5/O9-orientifold projection but it always breaks SUSY for the O3/O7-orientifold projection. The above action seems not immediately compatible with the way we identified the four dimensional $\N=1$ moduli: There is a $J$ instead of a $J\w J$ in the action. However, it was shown in \cite{Grimm:2004uq} that the $v^a=\int_{\Sigma^a_{(2)}}J$ defined in their equation (3.4) are functions of the moduli $\tau, G_a, T^a$ (see for example their equation (3.51) and the text below). Thus, we find that our approach works even for non-supersymmetric D5-branes in type IIB compactifications with an O3/O7 orientifold projection. Likewise we expect it to work for any other non-supersymmetric source, also if the source is not even a Dp-brane.

\section{Conclusion}\label{sec:conclusion}
In this paper we have studied string (flux) compactifications on $SU(3) \times SU(3)$-structure manifolds in which supersymmetry is broken by Dp-branes. We have shown that such setups \emph{cannot} be described using the standard language of linear, four-dimensional $\N=1$ supergravity but require the more general, so-called dS supergravity \cite{Antoniadis:2014oya,Dudas:2015eha, Bergshoeff:2015tra, Hasegawa:2015bza}. We have reviewed in explicit examples as well as in full generality that the bosonic contribution from the DBI action of non-supersymmetric Dp-branes to the closed string scalar potential takes a universal known form. In the case where we set the world volume scalars on the non-supersymmetric Dp-brane to zero we can easily include the new term in the scalar potential by modifying the K\"ahler and superpotential of the compactification in a universal way. All we needed for this is the fact that supersymmetry breaking Dp-branes have among their world volume fields the Goldstino, which is guaranteed by Goldstone's theorem \cite{Deser:1977uq, Cremmer:1983en,Grisaru:1982sr}. This leads to the very simple expression in equation \eqref{eq:KWmodified} that encodes the new contribution $V_{\rm new}$ from non-supersymmetric Dp-branes that includes the couplings of the non-supersymmetric brane to the closed string moduli.

Our general results should have important applications to flux compactifications and might help to improve our understanding of a tractable large class of non-supersymmetric string compactifications. It essentially allows one to include manifestly supersymmetry breaking terms to the scalar potential. For example, in the best studied class of type IIB flux compactifications with an O3/O7 orientifold projection, one can include supersymmetry breaking D5- and/or D7-branes. Both of these will lead to new terms in the scalar potential that are real functions of the closed string moduli and in particular of the K\"ahler moduli that usually do not appear at tree-level.

Extending our results to include all the open string world volume fields on a given non-supersymmetric Dp-brane in a particular compactification is an interesting exercise. So far, this has only been done explicitly for an anti-D3-brane in a warped throat in type IIB flux compactifications \cite{GarciadelMoral:2017vnz, Cribiori:2019hod}. However, while technically non-trivial there should be no conceptual problem in working this out for other cases. It might also be possible to find a more general way of including the open string moduli using the language of generalized geometry.

While we have focused here on compactifications to four dimensions that preserve $\N=1$ linear supersymmetry spontaneously broken by Dp-branes, many of our ideas apply much more broadly. For example, to compactifications to other dimensions and/or that preserve different amounts of supersymmetry. It should also not matter whether supersymmetry is spontaneously broken by Dp-branes or other sources like for example, M2-, M5-branes, NS5-branes or KK-monopoles. It would be interesting to study such related setups in more detail in the future.

\acknowledgments
We are thankful to F. Farakos, R. Kallosh, J. Moritz and E. Palti for useful discussions and E. Palti for bringing this interesting problem to our attention. The work of NC, CR and TW is supported by an FWF grant with the number P 30265. CR is furthermore supported by the Austrian Academy of Sciences and the Doctoral College Particles and Interactions with project number W1252-N27. The work of MT and AVP is supported in part by the KU Leuven C1 grant ZKD1118 C16/16/005. The work of MT is supported by the FWO odysseus grant G.0.E52.14N. TW thanks the KITP for hospitality and the organizers of the workshop ``The String Swampland and Quantum Gravity Constraints on Effective Theories'' for providing a stimulating environment during part of this work. This research was supported in part by the National Science Foundation under Grant No. NSF PHY-1748958.

\bibliographystyle{toine}
%\bibliography{refs}
\bibliography{supergravity}

\end{document}